\documentclass[twocolumn,prl,aps,showpacs,floatfix,superscriptaddress]{revtex4}

\usepackage{graphicx}

\begin{document}
\newcommand{\zrzn}{ZrZn$_2$}
\newcommand{\ZrZn}{ZrZn$_2$}
\newcommand{\uge}{UGe$_2$}

\renewcommand{\floatpagefraction}{0.5}

\title{Quantum Phase Transitions in the Itinerant Ferromagnet ZrZn$_2$}

\author{M. Uhlarz}

\affiliation{Physikalisches Institut, Universit\"at Karlsruhe,
Wolfgang-Gaede-Strasse 1, D-76128 Karlsruhe, Germany}

\author{C. Pfleiderer}

\affiliation{Physikalisches Institut, Universit\"at Karlsruhe,
Wolfgang-Gaede-Strasse 1, D-76128 Karlsruhe, Germany}

\author{S. M. Hayden}

\affiliation{H. H. Wills Physics Laboratory, University of
Bristol, Tyndall Avenue, Bristol, BS8 1TL, UK}

\date{\today}

\begin{abstract}
We report a study of the ferromagnetism of ZrZn$_{2}$, the most promising material to exhibit ferromagnetic quantum criticality, at low temperatures $T$ as function of pressure $p$.
We find that the ordered ferromagnetic moment disappears discontinuously at $p_c$=16.5~kbar.  
Thus a tricritical point separates a line of first order ferromagnetic transitions from second order (continuous) transitions at higher temperature.
We also identify two lines of transitions of the magnetisation isotherms up to 12\,T in the $p-T$ plane where the derivative of the magnetization changes rapidly.
These quantum phase transitions (QPT) establish a high sensitivity to local minima in the free energy in ZrZn$_{2}$, thus strongly suggesting that QPT in itinerant ferromagnets are always first order. 

\end{abstract}

\pacs{71.27.+a, 75.50.Cc, 74.70.Tx}

\vskip2pc

\maketitle


The transition of a ferromagnet to a paramagnet with increasing temperature is regarded as a canonical example of a continuous (second order) phase transition.  
This type of behavior has been well established in many materials ranging from nickel \cite{weiss26} to chromium tribromide \cite{ho69}.
The detailed variation of the order parameter near the critical point, in this case the Curie temperature, has been analyzed in a wide variety of systems using \emph{classical} statistical-mechanical models for the case when the Curie temperature is not too small. 
Classical statistics are appropriate when all fluctuating modes have energies much less than $k_BT_{c}$. 
%
%
It was pointed out by Hertz \cite{hertz76} that the system undergoes a \emph{`quantum'} phase transition (QPT) when the transition is driven by non-thermal fluctuations whose statistics are in the quantum limit.
%

The search for a second order (critical) QPT in itinerant electron systems, which are believed to be responsible for enigmatic quantum phases like magnetically mediated superconductivity and non-Fermi liquid behavior, has become of particular interest in recent years.
Experimental studies have thereby revealed notable differences from ``standard'' second order behaviour in \textit{all} materials investigated to date.
For example, in MnSi \cite{pfle97} and UGe$_2$ \cite{pfle02}, itinerant-electron magnetism disappears at a first order transition as pressure is applied.  
The bilayer ruthenate Sr$_3$Ru$_2$O$_7$, undergoes a field induced QPT with multiple first-order metamagnetic transitions \cite{perry04} and associated non-Fermi liquid behavior  in the resistivity \cite{grigera01}. 
However, these materials have complicating factors:  the zero-field ground state of MnSi is a helical spin spiral; UGe$_2$ is a strongly uniaxial (Ising) system; Sr$_3$Ru$_2$O$_7$ is a strongly two-dimensional metal.
In fact, theoretical studies suggest \cite{shimizu64,belitz99, vojta00, belitz02} that ferromagnetic transitions in clean three-dimensional (3$D$) itinerant ferromagnets at $T=0$ are \textit{always} first order.

In this Letter we address the nature of the ferromagnetic QPT experimentally.
The system we have chosen is the itinerant ferromagnet ZrZn$_2$, which is a straight forward itinerant ferromagnet with a cubic (C15) structure and small magnetic anisotropy.  
{\zrzn} has a small ordered moment ($M = 0.17 \mu_B$~f.u.$^{-1}$) which is an order of magnitude smaller than the fluctuating Curie-Weiss moment $\mu_{\mathrm{eff}}=1.9\,\mu_{B} \mathrm{f.u.}^{-1}$ and the Curie temperature is low ($T_c=28.5$~K).
The magnetisation is highly unsaturated as function of field up to 35\,T, the highest field studied.
Neutron diffraction in ZrZn$_{2}$ is consistent with all the hallmarks of a three-dimensional itinerant ferromagnet \cite{neut}.
Quantum oscillatory studies \cite{yate03} have shown that ZrZn$_2$ has a large quasiparticle mass enhancement \cite{yate03} as expected near quantum criticality.

Single crystals of ZrZn$_2$ have long been considered ideal in the search for quantum
criticality (second order behaviour).
Previous hydrostatic pressure studies suggested the existence of a second order QPT in ZrZn$_{2}$ \cite{smit71,hube75,gros95}. 
However, substantial differences of the critical pressure $p_{c}$ and the form of $T_{C}(p)$ were reported in different studies.
We now believe that these differences can be traced to low sample quality.
These studies underscore the need for high-quality single-crystals.

De Haas--van Alphen (dHvA) studies in high quality single crystals as function of pressure have recently  \cite{kimura04} let to the suggestion that multiple first order QPT exist in {\zrzn}, notably a crossover between two ferromagnetic phases at ambient pressure and a first order suppression of ferromagnetism at high pressure.
However, the evidence for a pressure induced QPT associated with the two ferromagnetic phases at ambient pressure was purely derived from a tiny pocket of the Fermi surface. 
Further, the first order QPT at $p_c$ was predicted theoretically but not experimental evidence has been reported until now.

Here we report an investigation of the question of multiple QPT in {\zrzn} using detailed measurements of the DC magnetisation, i.e., direct measurements of the order parameter. 
We establish for the first time that, while the ferromagnetic transition at ambient pressure is continuous, the ferromagnetism disappears in a first order fashion (discontinuously) as pressure is increased beyond $p_c=16.5$~kbar. 
The observation of a first order QPT at $p_c$ is strongly supported by the discovery of metamagnetic behavior, characterized by a sudden superlinear rise in the magnetisation as a function of applied field, for pressures above $p_c$.
We also characterize the pressure dependence of a second transition \textit{within} the ferromagnetic state for the first time using the DC magnetisation.
These data suggest another first order QPT.
Thus we establish experimentally the existence of multiple first order QPT based on measurements of the order parameter itself in the ideal candidate for the occurrence of ferromagnetic quantum criticality, {\zrzn}.

Two single crystals were studied, a short cylindrical piece and a half-cylinder, which produced identical results.
Data from these samples are therefore not distinguished further.
The samples studied here are the same in which superconductivity was originally discovered \cite{pfle01a}.
The method of growth avoids the problems with the zinc vapor pressure and is similar to that described in \cite{schr89}.
The residual resistivity ratio of our samples $\rho(293\,{\rm K})/\rho(T\to0) \approx 100$ is high and the residual resistivity $\rho_{0}\approx0.6\,\mu\Omega{\rm cm}$\ low.
Laue x-ray and neutron diffraction confirmed that the samples were single crystals.
Extensive dHvA data \cite{yate03}, the magnetic field dependence of the specific heat \cite{pfle01b}, $T_{C}(p)$ in very low fields for $p<16$\,kbar \cite{uhla02} and the electrical resistivity up to 21\,kbar and magnetic field up to 12\,T \cite{uhla04} of the same samples are reported elsewhere.

\begin{figure}
\includegraphics[width=.35\textwidth,clip=]{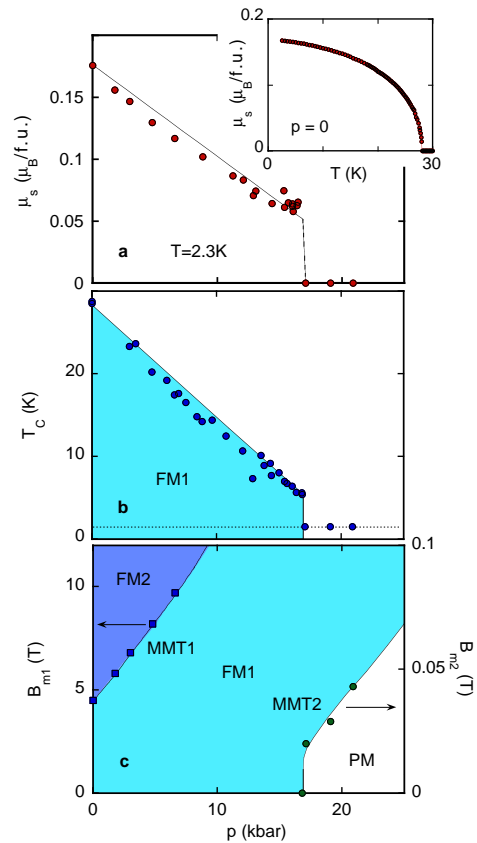}
\caption{(Color online) (a) Pressure dependence of the ordered magnetic moment.  The ordered moment was obtained be expolating magnetic isotherms (Arrot plots) to zero field. The inset shows the temperature dependence of the magnetization. (b) Curie temperature $T_{C}$ as a function of pressure. (c) Phase diagram determined in present measurements.  MMT1 ($H_{m1}$) corresponds to the `kink' or sudden change in gradient in the magnetisation isotherms previously observed in Ref.~\cite{pfle01a} for $p=0$. MMT2 ($H_{m2}$) corresponds a second kink observed for pressures $p>p_c$.}
\label{main-result}
\end{figure}

\begin{figure}
\includegraphics[width=.45\textwidth]{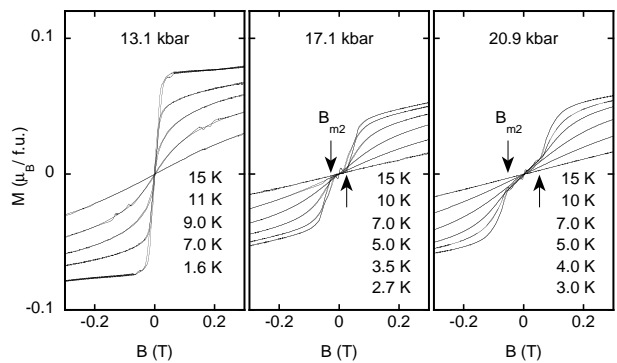}
\caption{Typical magnetization cycles below and above the critical pressure $p_{c}=16.5$\,kbar at various temperatures.  On the left hand side data correspond to the temperatures given in each panel, respectively.  The arrow marks the metamagnetic transition field $B_{m2}$ that appears above $p_{c}$.}
\label{hysteresis}
\end{figure}

The DC magnetization $M(B,T)$ was measured in an Oxford Instruments vibrating sample magnetometer (VSM) between room temperature and 1.5\,K at magnetic field in the range $\pm12$\,T.
Additional measurements were carried out in a bespoke SQUID
magnetometer in the range 4.2\,K to 60\,K at fields in the range 10
$\mu$T to 10\,mT.
The sample was measured together with the nonmagnetic miniature clamp
cell. 
The signal of the empty pressure cell was subtracted to obtain the
contribution of the sample, which was typically between 50 and 80\% of
the total signal. 
Pressures were determined from the superconducting
transition of Sn or Pb in the VSM and SQUID-magnetometer, respectively.

Fig.\,\ref{main-result}(a) shows the ferromagnetic ordered moment $M$ as a function at pressure and temperature (inset).  
The moment was obtained by extrapolating magnetization isotherms (Arrott plots) to zero field.
When the pressure is varied at low temperature ($T = $2.3\,K), the magnetization drops discontinuously at a critical pressure $p_c=16.5$\,kbar.
For comparison the inset shows the variation of the ferromagnetic moment at $p=0$ with increasing $T$ through the Curie temperature. 
At $p=0$ the transition is continuous (2nd order) presumably because we are in the classical (high temperature) limit.
In contrast, when the transition is suppressed to zero through the application of
hydrostatic pressure, the magnetization disappears discontinuously (first order). 
The Curie temperature $T_{C}$, shown in panel (b), qualitatively tracks $M$ and vanishes also discontinuously at $p_{c}$.

Fig.~\ref{hysteresis} shows magnetization curves near $p_{c}$.
Below $p_{c}$ (Fig.\,\ref{hysteresis}(a)) and at the lowest temperatures, $M(B)$ initially raises rapidly with $B$ as a single domain is formed.
Above $p_{c}$ a new feature appears at low fields where $M(B)$ initially increases approximately linearly. 
This is followed by a sudden superlinear increase with $B$ (`a kink'), at the field $B_{m2}$ marked by the arrows in Fig.\,\ref{hysteresis} (b) and (c). 
Above $B_{m2}$ the shape of the magnetization isotherms is reminiscent of those below $p_{c}$.  
A sudden rise in $M(B)$, such as that observed here, is usually called metamagnetism \cite{wohlfarth62}.
It is a deep signature of a first order structure of the underlying free energy (local minimum) and proofs unambiguously that the QPT at $p_c$ is first order.
The magnetization isotherms allow us to extract the pressure dependence of the cross-over or metamagnetic field $B_{m2}$ as shown in Fig.\,\ref{main-result}(c). 
Our data are consistent with $B_{m2}$ terminating near $p_{c}$ showing an intimate connection with the discontinuous drop at $p_c$.

\begin{figure}
\includegraphics[width=.35\textwidth]{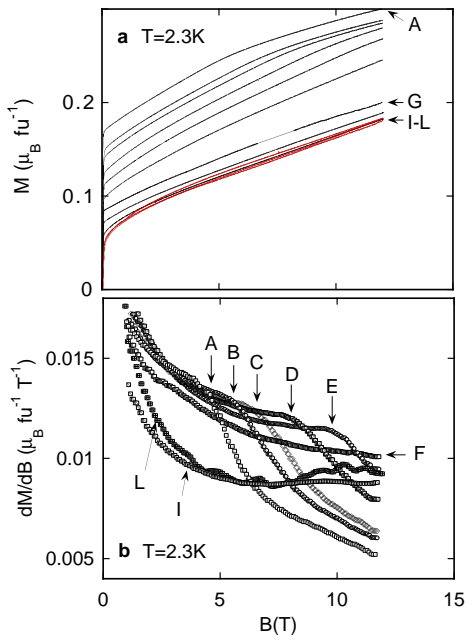}
\caption{(Color online) (a) Magnetization $M$ as function of magnetic field $B$ at $T=2.3$\,K as function of pressure.  Labels correspond to the following pressures in kbar: A=0, B=1.8, C=3.0, D=4.8, E=6.6, F=8.9, G=12.5, H=13.1, I=14.4, J=17.1, K=19.1, L=20.9.  (b) Derivative $dM/dB$ for selected pressures (a).  For pressures A-E the arrow marks the crossover field $B_{m1}$.}
\label{magnetization}
\end{figure}

In addition to the low-field magnetization measurements described above, we also made measurements at high fields under hydrostatic pressure. 
Fig.\,\ref{magnetization}(a) shows the field dependence of the magnetisation for various pressures up to 20.9~kbar. 
As with our previous study at ambient pressure \cite{pfle01a} we observe a kink in the magnetization near $B \approx$\ 5~Tesla (curve A in Fig.~\ref{magnetization}(a)). 
The anomaly can be seen more clearly in the derivative $dM/dB$ shown in Fig.\,\ref{magnetization}(b). 
We are able to identify a cross-over field $B_{m1}$ from a low-field phase (FM1) to a high-field field phase (FM2), where the ordered moment of the high field phase is increased by $\sim$10\% (see e.g. Ref.\,\cite{kimura04}). 
With increasing pressure $B_{m1}$ increases as plotted in Fig.\,\ref{main-result} (c). 
The transition from FM1 to FM2 at $B_{m1}$ corresponds to a transition \emph{within} the ferromagnetic state.
By analogy with $B_{m2}(p)$ we extrapolate that $B_{m1}(p)$ terminates at a QPT at approximately -6~kbar.

\begin{figure}
\includegraphics[width=.45\textwidth]{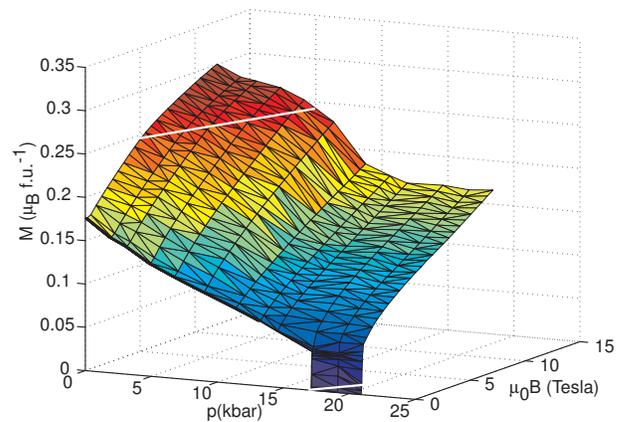}
\caption{(Color online) The experimental variation of the magnetization $M$ with pressure
and applied field in ZrZn$_{2}$.  The Figure is based on the data shown in Figs. \ref{hysteresis} and \ref{magnetization}.  The white lines show approximately the locations of `kinks' in the magnetization reported in this paper.}
\label{M_color}
\end{figure}

Fig.~\ref{M_color} shows an overall representation of the magnetization based on the data
in Figs.~\ref{hysteresis} and \ref{magnetization}. 
The white lines denote the approximate positions of the transition fields $B_{m1}$ and $B_{m2}$.  
It is interesting to note that $B_{m1}$ occurs at approximately constant $M$.
This strongly suggests that the transition is triggered by an exchange splitting that is insensitive to pressure and therefore indeed related to the electronic structure.
As for the transition at $B_{m2}$ the anomaly at $B_{m1}$ hence is evidence of a further first order minimum in the free energy, establishing that the associated QPT must be first order.
Fig.~\ref{M_color} also shows that the high field magnetization varies very little with pressure above the critical pressure $p_c$.

\begin{figure}
\includegraphics[width=.30\textwidth]{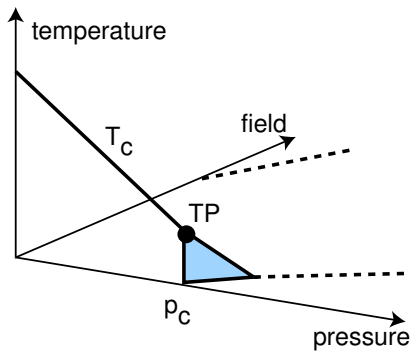}
\caption{(Color online) Systematic representation of a possible phase diagram of ZrZn$_2$ proposed in Ref.~\cite{kimura04}. For $p \leq p_c$, a line of second order ferromagnetic ends at a tricrital point TP. For $p > p_c$, a first order jump in the magnetization occurs on crossing the shaded (blue) area which extends as a crossover to higher pressures (dotted line).  At higher fields a crossover in $M(B)$ persists which emerges from a QPT at an extrapolated negative pressure.}
\label{f:tricritical}
\end{figure}

We now discuss the interpretation of our results.
Our observations show for the first time that the ordered magnetic moment in ZrZn$_2$ disappears discontinuously around $p_c \approx$\ 16.5~kbar.
This suggests that in the $p-T$ plane, a tricritical point separates a line of first order transitions from second order behavior at high temperature.  
Fig.~\ref{f:tricritical} shows the proposed schematic phase diagram for ZrZn$_2$  \cite{kimura04}, which has not been verified experimentally until now. 
Crossing the shaded (blue) region corresponds to a first order phase transition.  
The related dotted line represents a crossover where there is a rapid change in $M(B)$. 
In our experiments we observe a discontinuous drop of $T_c(p)$, consistent with a tricritical point near $T_{t} \approx 5$~K and $p_{t} \approx 16.5$~kbar as qualitatively proposed in \cite{kimura04}. 
However, we have not observed a discontinuous change of $M$ with $T$ or $B$. 
Presumably this is because the region of first order transitions is small and none of the pressures chosen in our study sample this region.  
At high magnetic fields (second dotted line) we observe an unusual sudden change of the gradient in the magnetization isotherms $M(B)$ that translates into an increase of the ordered moment.

The evidence for first order behaviour we observe bears on various theoretical descriptions of ferromagnetic phase transition in metals in the quantum limit.  
The first is a `Stoner' picture where the effect of electrons are incorporated into a one-particle band structure and the exchange interaction is described by a molecular field $\lambda M$.  
In this case, the condition for ferromagnetism and field dependence of the magnetization $M(B)$ are determined by the structure of the electronic density of states near the Fermi energy.  
Within this model, Shimizu \cite{shimizu64} has shown that, if the Fermi energy lies near
a peak in the density of states, ferromagnetism will disappear at a first order transition.
An extension to this model \cite{wohlfarth62,Sandeman03} suggests that the application of a magnetic field can lead to a metamagnetic QPT, if the criterium for ferromagnetism is not quite satisfied. 
The applicability of the `Stoner' model to ZrZn$_2$ is supported by band structure calculations \cite{santi01,singh02} and the experimental determination of the Fermi surface \cite{yate03}, which suggest that the paramagnetic Fermi energy lies about 30~meV below a double-peak in the one-electron density of states.
In a second description of ferromagnetic quantum criticality \cite{belitz99,belitz02},
the transition is found to be generically first order due to a coupling of long-wavelength magnetisation modes to soft particle-hole excitations.
Near the ferromagnetic QPT, this leads to a nonanalytic term in the free energy that generates first order behaviour.  
This mechanism is independent of the band structure and \textit{always} present.

In summary,  we report for the first time that ferromagnetism in {\zrzn} disappears discontinuously at $p_c=$\ 16.5~kbar and `crossover' or `transition' lines exist in the $p-B$ plane ($B_{m1}(p)$ and $B_{m2}(p)$).
The first transition, $B_{m1}(p)$, occurs at ambient pressure and its pressure dependence shows the existence of a first order QPT at negative pressure.
The second transition, $B_{m2}(p)$, is associated with the disappearance of ferromagnetism at the first order ferromagnetic QPT at $p_c$.
Our experiments establish for the first time the existence of multiple first order QPT in {\zrzn}, as previously proposed \cite{kimura04}.
The emergence of these multiple first order QPT with decreasing temperature in a material, {\zrzn}, that is in every respect considered to be \textit{the} prime candidate for ferromagnetic quantum criticality supplies strong evidence that QPT in itinerant ferromagnets must be generically first order.

We wish to thank D. Belitz, S. Dugdale, M. Garst, J. K\"ubler, H. v.  L\"ohneysen,
G. Lonzarich, I. Mazin, K. Noriyaki, A. Rosch, G. Santi, T. Vojta and M. Vojta.
Help by I. Walter, T. Wolf, V. Ziebat and P. Pfundstein are
gratefully acknowledged.
Financial support by the Deutsche Forschungsgemeinschaft (DFG),
European Science Foundation under FERLIN and the UK EPSRC are
gratefully acknowledged.

\end{document}